\documentclass[]{article}

\usepackage[utf8]{inputenc}

\usepackage{fullpage}

\usepackage{amsmath}
\usepackage{amssymb}
\usepackage{amsthm}
\usepackage{latexsym}
\usepackage[font=small,labelfont=bf,labelsep=space]{caption}
\usepackage{float}
\newtheorem{definition}{Definition}

\usepackage{boxedminipage}

\usepackage{listings}

\usepackage{minitoc}

\usepackage{ifpdf}

\ifpdf
\usepackage[pdftex]{graphicx}
\else
\usepackage{graphicx}
\fi
\title{Shaping ideal cities: the graph representation of the urban utopia}
\author{Roberto D'Autilia\\
{\sl Dipartmento di Architettura} \\
{\sl University Roma Tre}\\ 
{\sl Rome, Italy}\\
{\sl roberto.dautilia@uniroma3.it}\\\\
Marco Spada\\
{\sl Department of Geography and Planning}\\
{University of Liverpool}\\
{Liverpool, United Kingdom}\\
{\sl marco.spada@liverpool.ac.uk}
}

\begin{document}

\ifpdf
\DeclareGraphicsExtensions{.pdf, .jpg, .tif}
\else
\DeclareGraphicsExtensions{.eps, .jpg}
\fi

\maketitle

\begin{abstract}
The ideal Renaissance city is designed as a star-shaped fortress, where the streets and squares are organized to speed the movement of people and soldiers.
Symmetry and accessibility represent the key features for the organization of the urban space.
The resulting city is hierarchized and does not always guarantee an optimal degree of connectivity.
Taking the baton from the work done by space syntax in the definition of properties of spatial graph representation, we introduce a method to compute urban graphs from the Euclidean representation, the corresponding line graph and the contraction of nodes with the same urban function.
We analyze the urban graphs of five historic cities: Vitry le Fran\c{c}ois, Avola, Neuf Brisach, Grammichele and  Palmanova and compare the analysis restults with the corresponding results from space syntax.
Analysis of the spectral gap and the relative asymmetry distribution show a similar structure for these cities.
The irregular or reticular housing structure seems to ensure connectivity and accessibility more than the regular grids.
However connectivity is ensured by the most peripheral streets, which in the space syntax representation play a marginal role.
\end{abstract}
\pagebreak
\section{Introduction}

The idea of city has undergone, over time, a slow transformation reflecting the way we design, build and live the urban fabric.
The primitive city, the space that starts when ``the trails become roads'' \cite{Lelievre}, was little more than a cluster of houses within a defensive palisade; the founding rituals of the city, however, established its sacredness, and the rituals that made the agglomerate an inherently sacred space made it possible to locate, in the core of the city, spaces for pastoral rituals, of primitive derivation, and for ceremonial games. 

The urban space assumed different meanings according to different founders and different cultures.
The oikistic literature set in the late Roman period, from Plutarch to Procopius of Caesarea, was intended to compare the space of the sacred cities of antiquity with that of secularized cities of their contemporaneity, in which the concept of city was increasingly as that of ``citizenship''.

The crime of ``usurpation of citizenship'', defined as the system of laws and rights which Roman citizens could enjoy within the state territory, was punished with death, and the granting of citizenship was used through numerous laws by  emperors as instrument of loyalty-building of war allies.
The extension of the rules on citizenship, along with the increasingly absolutist control by the emperor, led primarily to a substantial uselessness of the institution-privilege of ``citizenship'' and the impoverishment of its archaic and religious meanings, and in the second instance contributed significantly to the crisis of the empire in the fourth century.

From the city made up of men narrated by Thucydides \cite{2009peloponnesian} we come to the city made up of streets, more and more assimilating town planning to the description of places connected by roads, through the removal of the ritual identification process of the house as a city and vice versa.
The phenomenon is in itself conceivable, in cities in which we discuss the ``right to mobility'' to reach the workplace or place of leisure and rest; the street network takes a different value than the more properly ritual and cultural urban facts.

Joseph Rykwert \cite{rykwert1988idea}, following the mentioned oikistic literature, emphasizes how the construction of a city is not only an expression of urgent military or economic needs, but is also the result of cultural overlays of civilization that proposed an idea of city even before the first hut was built.
Secularization of civilization occurred between the seventeenth and nineteenth centuries, and the consequent definition of the Industrial Revolution \cite{rykwert1988idea} has, first, freed architectural expressiveness towards the search for new forms and materials, and secondly brought to the city huge masses of people, inventing a new kind of city, the ``periphery'', which until that time simply had not been contemplated in urban culture, and shifting the focus of the city planner from the elaboration of an iconographic system of the city, to the creation of infrastructure useful to prevent the outbreak of pestilences in densely populated areas \cite{benevolo1977history}.

The demolition of the city walls has been one of the symbols of this secularization.
Sections of walls considered sacred until, de facto, the Thirty Years War, were suddenly needless, due to changes in the theory and technology of warfare.
Within a few years the sacredness of the border, prescribed by Leo III the Isaurian in the Agrarian Laws \cite{d2014estimo} and by Rotari with his Edict, was lost.

Construction of major boulevards where the milestones of Pomeria had stood in Roman times had the effect of opening to sprawl outwards and, conversely, redefining the historical center as unhealthy and decadent.
Paris and Vienna have experienced this development toward the outside, and even inside the walls, upwards, in a few years.
Rome, thanks to a substantial inviolability of the Aurelian walls, underwent development outside of the walls only in the 1920s development regulated approximately with the General Variation to the Plan of 1909, 1924, and finally with the 1931 Plan.

It is in the Renaissance that the city became the center of a wider discourse that concerns the measure of man, the calculation of the quantities of the universe in relation to the time of man, to the human dimension in relation, finally, to the processing and production of human ideas.
As the grid by Hippodamus of Miletus shows, in planimetric terms, the democratic culture of the pre-Hellenistic Greek polis, and the Roman reticulum follows the lead of rural ritual city-founding acts, processed subsequently in a syncretic way with military rituals, to make the city of citizens a sort of stone-and-wood military  camp, so the Renaissance city exceeds the walls of the medieval settlement within which it has been conceived and develops new forms for the walls, the dwellings, the streets, and public and religious buildings.

It is important to point out that the experience of the cities of the Renaissance, although extremely secularized from the point of view of the production of building typologies, remained strongly anchored to a Christian-Catholic ideal of the city.
The central pavilion of the ``Tavola di Urbino'' is topped by a cross, while the late Romanesque church is located off-axis with respect to the urban structure; the derivation, however remains Roman: still, in the Urbino panel \cite{krautheimer1948tragic} the perspective grid is calculated on the basis of a 73 mm square that determines the space of the {\sl piazza} almost the same size, divided by 10, of the Vitruvian ``passus minor'', or  Gressus, equal to 2.50 Roman feet (about 74 cm.). 

The Urbino panel, therefore, is a reduction in scale of a  hypothetically existing city, measurable and estimabile from its general features (spaces, volumes and dimensions) its particular ones (structures, windows, doors, thresholds, marbles). This is therefore a real city in its two-dimensional interpretation. The same can be said of the Baltimore panel, in which the composition, alone among the three panels, illustrates a real space of the piazza, slightly recessed with respect to the plane of the architectural monuments and defined by the fountain and by the commemorative columns.

Obviously, the ideal city had become a city ideally formed on the basis of secure mathematical measures, and not ``the ideal city to live in'': to imagine a city where the design of the houses and buildings takes into account the movement of the sun, the force of the winds, the natural slope of the land is far removed from the ``idealism'' of Renaissance urban morphology.
The ideality in this case is to be found instead in the geometry of the measure, the possibility of iteration or completeness of orthogonal or radial grids, in the urban role of the walls, in the hidden geometry of relations between built and unbuilt spaces.

At first glance the urban fabrics of the cities that we will analyze are essentially repetitions of a common morpho-typological trait, mirrored or moved in a substantially symmetrical way.
Closer examination, however, shows this is not so: symmetry, often abhorred as a trivializing of architectural complexity, is constantly denied by the exceptional building, by the urban out-of-range, by the the mass of the church or of the civic building.
In other words the ideal cities, or even, as we shall see, the city-fortresses, conceal their complexity in their asymmetry, which, far from egalitarian trivializations, automatically designates taxonomies of spaces and hierarchies. 

The cultural base of the ideal city is deeply syncretic: it derives from the interpretation of the Roman fabric carried out by men of the Middle Ages, steeped in Christian culture and projected toward an uncertain future.
The origin of these cities, whose historical development will continue until the eighteenth century, is almost always traumatic: either war or destruction due to natural events.
In other words, the Berlin, Baltimore and Urbino images, topoi of serenity and contemplation, have become houses and streets just to oversee a territory or regain it after wave of destruction.

In this paper we suggest a method of processing of the urban morphologic fabric: from the urban grid we extract the non-directed graph which identifies the relationships between the individual sections of public space.
This graph is processed according to an elementary taxonomy  which locates in the urban fabric a number of invariant features; finally, the relationships between the different taxonomy entries, are studied based on analytical considerations, in order to identify the topology between the individual parts that underlies the urban structure.

Graph theory is a powerful tool for analyzing city usability and structure.
Originally introduced by the Swiss mathematician Leonhard Euler to solve the K\"onigsberg bridges problem \cite{Alexanderson_AMS2006}, it is now a widespread means for the study of urban topologies \cite{Diestel2005} \cite{BV_Buch}.

The methods and results of the space syntax \cite{Hillier1999} are in fact based on a computer implementation of some graph analysis algorithms which make evident such urban features as the centrality, the closeness and the betweenness of the open spaces.
In addition, the comparison of the graphs derived from the real-world with the theoretical ones (k-graphs, paths, complete graphs) provides a deep insight into how real cities are organized and how much they differ from their possible ideal shapes.

The theoretical urban morphologies are, in general, the result of a tradeoff, and this encourages one to investigate how much the real cities deviate from the idealized shapes \cite{BV_Buch}.
We consider the contribution of self-organization \cite{BarthelemyBBG13} to the improvement of the connectivity and the accessibility of the city, and examine whether prescriptive planning is more useful for the infrastructures than for the growth of the living spaces.

To give an example, in 1852 Baron Haussmann substantially modified the urban structure of Paris by {\sl building safer streets, large avenues connected to the new train stations, central or symbolic squares and improving the traffic flow and the circulation of army troops} \cite{BarthelemyBBG13}.
The action of Haussmann was essentially a reshaping of infrastructures, where some local growth processes aggregated on these structures, contributing to the overall accessibility of the city.

The Haussmann's intervention was not the first of its kind.
The Ippodamus grid \cite{1997politics} and the circular Iranian cities \cite{Mohajeri201210} \cite{karimian2010preliminary} are very ancient examples of this way of {\sl a priori} urban thinking.
Althougth it was useful for infrastructures optimization, the abstract view charged the city shape with political and rhetorical issues, putting on two opposite sides the ``imagined city'' and natural or spontaneous growth, conditioned by demographic contractions and expansions and by other social and economic occurrences. 

\begin{figure}[ht]
\centering
\includegraphics[width=140mm]{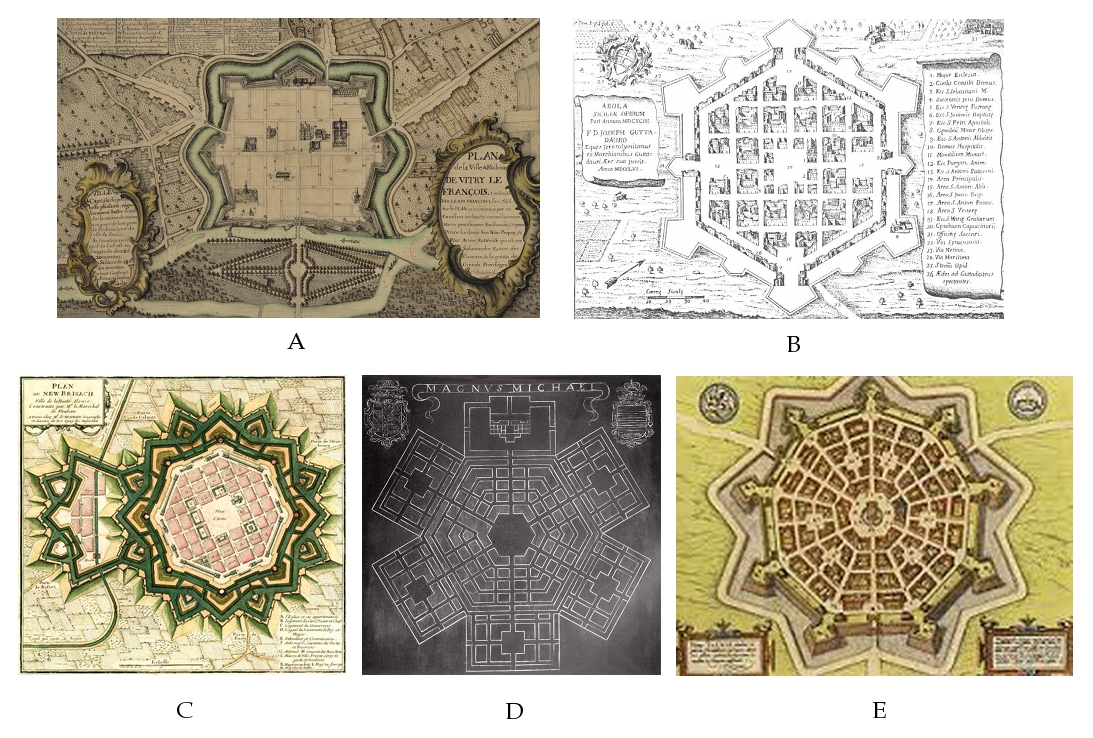}
\caption{The ideal cities we analyze: (A) Vitry le Fran\c{c}ois, (B) Avola, (C) Neuf Brisach, (D) Grammichele (E) Palmanova}
\label{lecitta}
\end{figure}

In this framework we study the structures of five ``ideal cities'', five small examples of what happened a few centuries later in Paris, and compare the planned city with its inner spontaneous evolution.
In particular, we examine the urban ideal morphologies during the historical period between the construction of the French city of Vitry le Fran\c{c}ois (1545) and the reconstruction of the city of  Neuf-Brisach (1698) in Alsace, on the border with Germany.
During that time, the Italian city of Palmanova (1593) was also built and, after the earthquake of 1693, the cities of Avola and Grammichele were reshaped, according to new political and military criteria, by Italian and Spanish-Dutch engineers (Fig.\ref{lecitta}).

We chose these five cities because they show a simple geometry, but represent also a sort of general outline one can use to understand the relationship between structural and infrastructural shapes.
However we point out that it is not possible to make a clear distinction between these two features for open spaces.
For example the inner courtyards of the ancient Ostia's insulae had a function of internal facility \cite{propylaeumdok516}, but in a short time some of these courtyards assumed the function of connecting external streets, as though they were public squares \cite{kaiser2011roman}.

Within this scheme we identify some urban features of the ideal city which affect the flows and the accessibility of the urban environment and analyze them in the framework of graph theory.
To this purpose however, we need to generalize the space syntax dual construction in the way suggested by  Sergio Porta, Paolo Crucitti, and Vito Latora \cite{Porta2006} to guarantee the continuity of the infrastructural lines (see methods section).
Our method, based on labelled {\sl line graphs} \cite{Diestel2005}, is quite general and can be exploited to include non-spatial measurable properties of the urban spaces, as for example the acoustic properties of the streets.

The analysis shows a structural homogeneity for the five cities, but evidences also a connectivity determined by the streets along the external walls.
The properties of the spatial graph of the ideal cities also indicate that there were no areas of urban segregation, and finally, we found that even when the city is connected and accessible, the overlying star-shaped structure makes it vulnerable when it is not broken by a small amount of spontaneous and desordered growth.

The paper is organized as follows: in Sec.\ref{idealcity} we will show how a rethorical construction of the space of the ideal city could be abstracted and analyzed in a set of qualitative urban measures. This part represents a theoretical approach in which a cultural and symbolic background become architecture and urban planning; as we will see, starting from a historical-critical approach we could underline a set of uniform features, commons to all the case studies. In Sec.\ref{GU} we will highlight the idea of an urban graph, extrapolated by former taxonomy, and we will draw the method of labeling and contraction of the urban graph, as explained in Sec.\ref{idealcity}. In Sec.\ref{results} we will show the results of the modeling process and the differences between the cities.  Finally, in Sec.\ref{conclusions}, we will conduct a critical discussion of our proposed model.

\section{The Ideal City}
\label{idealcity}
The idea of the perfect city arose with the definition of the city as a place of democracy and emancipation from the slavery of rural life.
{\sl It is a common conviction, amongst urban practitioners, that the spatial form of a city expresses, in a more or less direct way, the social form of its community.
So strong and old is this belief that a significant number of ideal cities have been designed, throughout the ages, in such a way so as to prefigure the ideal society of individuals that would have lived in them} \cite{Piccinato2002}.		

The Italian Renaissance expresses a need for a political, military and also conceptual presidium of rural areas symbolized by geometric cities.
The will to create poles for the urban middle class together with the need to defend them, generated the ``fortress city'' of Northern Italy, among which Palmanova may be considered the most representative.

The Renaissance ideal city, revised in the following centuries, was theorized by Filarete (Antonio Averlino). He drew up the theoretical city of Sforzinda in 1460, subsequently the subject of discussion by theorists, artists and architects, among them Fra Giocondo, Daniele Barbaro, Giorgio Vasari and Giorgio Martini \cite{rosenau1974ideal}.
Interest in the ideal city was also influenced by military engineering, which ensured the safety of the city in a historical period characterized by frequent local conflicts.
In Europe there are several examples of military citadels built on pre-existing medieval settlements and few examples of new settlements.

We said earlier that the nature of the ideal city is strongly syncretic.
On the one hand the secular attempt to create a city far from the chaos of the historically established cities, on the other a missionary dimension of rebuilding of the Heavenly (or New) Jerusalem, have shaped the ideal city as a complex system of meanings, in which secularism and Christianity merge into a unique layout.
One of the models for the ideal city, therefore, is the Heavenly Jerusalem described by John in the Book of Revelations, first description in urban terms of a highly stratified concept in early Judaism never described before.
The city of John is a transformation of the Holy of Holies of the Temple of Jerusalem: a cubic buiding with in the middle the Tree of Life and three doors on each side.
This description, which in part determined the fortune of the idea of the Renaissance ``Roma Quadrata'' (Square Rome) as a Christian alternative to Jerusalem, has developed a model in which to the elements constituting the ideal city are given a number of symbolic or metaphorical meanings.
The gate, which in our system of labels we define by the letter $(G)$, already has a symbolic role in the founding of the city.
There is an entire chapter of Rykwert \cite{rykwert1988idea} that focuses on guarding gods of the gates and the role of the founder which, according to the Etruscan rite, for three times raises the plow in correspondence with the gate space, but it is with Christianity, founded on the gospel of John, that the gate is given a value of liturgical and salvific nature.
John writes (10:9, KJV) {\sl I am the door: by me if any man enter in, he shall be saved, and shall go in and out, and find pasture.}

In this sense we can gather under a single heading of ``ideal city'' a relatively large number of cities, from those more properly fortified, the cities-fortress, where we saw that a special role is given when founding by the priest-general, to the polygonal cities, not military, in which the founder gets help from a haruspex-priest for the sanctification of the city: this the case of the Sicilian cities destroyed and rebuilt after 1693, when a Jesuit monk, Angelo Italia, collaborated on the design of all the great reconstructed cities, including Avola and Grammichele, studied in this paper.

It makes little sense, therefore, to distinguish between cities, as the symbolic and allegorical nature of all responds substantially to that syncretic model we mentioned above, in which the Celestial Jerusalem is added to the cardo-decumanic system of the city founded by the Romans; even in the ideal city there is a cardo, in the system of labels $(C)$, to which correspond one or more decumani, in the system of labels $(D)$. As we can see in the old Etruscan cities, like Kainua-Marzabotto, several decumani may correspond to a cardo.

This step should not be underestimated.
The technique of dividing land into equal parts, and this accuracy into the ideal cities is extreme.
It was known, in Roman times, only to a small number of technicians, ``gromatici'', organized into a kind of ritual brotherhood, and their technique of land surveying, the ``Agrimensura'', Varro recalls, was derived from Etruscan knowledge.
In this sense the ideal city, and its urban components, play a role of pre-Roman rituality that, in some way, was handed down to Italy in the sixteenth and seventeenth centuries.
To a more complex tradition, instead, one can connect the role of the central square, which in the system of labels we will call $(S)$: a location certainly functional for assemblies and for markets, the square focuses lots of other features and symbols, too.

Rykwert \cite{rykwert1988idea} always emphasizes that, in Kainua, at the intersection of the cardo and decumanus, was placed a memorial stone with a cross mark, probably to indicate the point where the surveyor had driven the groma, which is cross-shaped.
This place, therefore, would indicate a substantial {\sl umbilicus mundi} of the city, present in ancient times, and still not fully understood in its historical and symbolic implications.
At the same time we observe that the description of the Heavenly Jerusalem put the tree of life right in the city center, in a space that apparently seems to be a huge square. Once again the Etruscan-Roman-Christian syncretism becomes urban matter.

The construction of walls generates an extremely complex space, especially in the construction of a polygonal city, in which the walls cast shadows, make the city unhealthy and define the streets and public spaces.
The furrow traced by the founder, as narrated by Plutarch, generates, because of the ritual and sacred nature of the act of plowing the city, three spaces, one inside, holy, where the sod raised by the plow is laid, one outside, spirits and ghosts are confined, and the furrow itself, the sulcus primigenius, on which the walls would have been erected.
The Renaissance city is not, however, the Roman city, and, whereas in the Roman town houses could be built against the inner side of the walls, immediately after the capture of Jerusalem during the Fourth Crusade of 1204 it was clear that, for military purposes, it was preferable to leave a large space between the walls and the city.


The two roads that run along the walls, inside and outside, are very different entities, from the ritual and urban standpoints.
The outer road, in the labels $(E)$, is a ritual space and pomerium, in which resides something other with respect to the sacredness of the city, and the inner one $(M)$ is a military control space.

Finally, we must analyze the internal structure of the city, i.e. the grid of streets, in the system of labels $(A)$ which, by drawing the space, determines the shape of the buildings.
Basically we have two types of agglomeration, the first Hippodamian, the second radial.
In these cases, contrary to what can be observed in Etruscan and Roman city, there is not a symbolic or ritual specificity that determines one or the other dimension; moreover, these cities being often reconstructions, it can be assumed that the choice of a fabric does not depend on decisions related to the previous configuration, in the case the rebuilt city, but on more functionalist determinations (probably of a military nature and/or linked to the nature of the soils).

In this space, because of the shape of the roads, alleys and small squares may be generated.
In our analysis we have interpreted these spaces as streets, because our analysis has tagged the sites with respect to their adherence to a principle of symbolic interpretation. Of course, and here lies the deep and critical self-determination of the architect and of the urban planner, if one wants to make a graph based on the land use, or on the public or semi-public nature of the spaces, or - for example - on the level of noise of a place with respect to another, naturally the graph would be different. 

In this paper we are simply disseminating a method of analysis that is suitable for any personal consideration on urban space. We are aware of the limits of some highly automated quantitative methods, which invalidate the critical reading of the space through the architect's eye, and we strongly believe that the labeling system, which is given here as an example, should be rewritten and redesigned in every study.

Some invariants, with minor differences, can be identified in the structure of the ideal fortress-city.
These invariants are found both in military installations and in civilian infrastructures.
These components are found in every ideal city, with few differences.
As shown in the next section, we identify these invariants on the spatial graph by means of a set of labels.

\begin{center}
	\begin{table}[h]
    \begin{tabular}{   c | l }
    Label & Description \\ 	[.1cm]\hline \\ [-.3cm]
    $C$ & Cardus  \\ 
    $D$ & Decumanus. Sometimes the decumanus is replaced by two diagonal streets. \\ 
    $S$ & The main square, always at the centre of the city where the main streets converge. \\ 
	$G$ & The city gates on the external polygonal walls. \\ 
	$E$ & The outside ways that run parallel to the walls. \\ 
	$M$ & The inside ways that run parallel to the walls. \\ 
	$A$ & The grid of the houses, sometimes with non-ordered structure. \\ 
    \end{tabular}
	\caption{The labels system for the structural elements of the ideal city.}
	\label{etichette}
	\end{table}
\end{center}

\section{The Urban Graphs}
\label{GU}
Space syntax is a method for analyzing the accessibility and other spatial features of the cities.
The definition given by Hillier, Hanson and Graham in 1987 \cite{RePEc:pio:envirb:v:14:y:1987:i:4:p:363-385}, reported in \cite{Ratti2004}, highlights how the theory is founded on the spatial configurations of the city, but rightly does not establish any representative quantity for these configurations:\\\\
{\sl“Space syntax (...) is a set of techniques for the representation, quantification, and interpretation of spatial configuration in buildings and settlements. Configuration is defined in general as, at least, the relation between two spaces taking into account a third, and, at most, as the relations among spaces in a complex taking into account all other spaces in the complex. Spatial configuration is thus a more complex idea than spatial relation, which need invoke no more than a pair of related spaces.”}\\\\
Starting from these ideas \cite{HillierSLS}, the theory has been widely extended and applied to different case studies, stimulating also the development of computer programs to implement the methods \cite{VaroudisX}.
The main variables of space syntax are the axial lines derived from the isovists which identify the open spaces \cite{RePEc:pio:envirb:v:34:y:2007:i:3:p:539-555}.
The interaction between two spaces or the corresponding axial lines is given by their intersection, resulting in a direct connection of the corresponding spaces and the possibility to move from one to the other.

The model gave fruitful results and an original point of view on the structure and the function of a city \cite{taus}\cite{RePEc:pio:envirb:v:34:y:2007:i:3:p:539-555}\cite{PhysRevE.73.066107}, but also gave rise to discussion about its generality and possible incongruences \cite{Ratti2004}\cite{RePEc:pio:envirb:v:31:y:2004:i:4:p:501-511}\cite{RePEc:pio:envirb:v:31:y:2004:i:4:p:513-516}.
On the mathematical side it has been proven that if we associate a node with each axial line and connect two nodes when the corresponding lines intersect, the space syntax can be reformulated in terms of graph theory \cite{BV_Buch}.
It has been shown that such analysis is linked to the theory of the stochastic processes, and in particular to the stationary measures of random walks on a graph \cite{hillierparadigm1999}.
Numerical tools have been developed to identify the graphs from open spaces, axial lines and their intersections \cite{VaroudisX}.

Historically space syntax focused mainly on the visual experience.
However urban spaces can be perceived also by other senses, and in general their characteristics can be measured by observations and instruments.
A person who moves in the city also perceives it as sound, odour, crowd, danger, cultural references or, in general, as the result of the behavior of variables not immediately identifiable by the view.

The visual aspect is certainly dominant, but the spatial view is just one of the possible variables that identify the spaces.
In fact cities do not connect only open spaces, but also connect services, social behaviors, emergencies and general urban quantities.
It is therefore necessary to define more general urban variables then the axial lines, extending the space syntax to include also quantities which are not simply visual.
Indeed the city is a structure connecting all the urban features, not only the open spaces \cite{Batty:2013:NSC:2553103}.

If we define a primary graph $G_p$ on a city map by the identification of the vertices with the road crossings and the edges with the streets and the squares, then the axial graph is related to the corresponding line graph $L(G_p)$, the graph on the edges where two nodes are connected if the corresponding edges of the primary graph are adjacent \cite{Diestel2005}.
By assuming this point of view, any property that can be directly measured in an open space is a variable defined on the edges of the primary graph.
It is consistent the literature about the relationships between measures like crime, services, housing density or real estate prices and urban networks \cite{DaviesRiots2013}\cite{gauvin:modeling}.
It is therefore natural to look for a rigorous process to get the analysis of the space syntax starting only from the primary graph and exploiting the properties of the corresponding line graph.
To this purpose different proposals have been made, dealing mainly with the structural continuity between adjacent spaces.
The proposals range from the measurement of the angles between the axes \cite{RePEc:pio:envirb:v:34:y:2007:i:3:p:539-555} to the introduction of properties of the edges of such as, for example, the street names \cite{Porta2006}.

We introduce a method to generalize these proposals \cite{crucitti2006centrality} as well as to clear up the terminology.
In fact the term ``dual graph'' is used to identify the axial graphs of the space syntax \cite{BV_Buch}, where in the language of graph theory the duality indicates a different graph transformation \cite{Diestel2005}.
Also, unlike the spatial duality, the axial graphs are not involutive because the dual of the dual of the axial graph is not the original one.
The method exploits the primary graph, the corresponding line graph and the graph contraction algorithms \cite{dautiliaMathematica2015}, giving as a particular case the results of the space syntax.

If the primary graph $G_p$ is the graph where the edges are the open spaces of the urban map and the nodes are the crossings between them, then to each node of $G_p$ some spatial coordinates are also associated.
A street is an edge of $G_p$ and the intersections with other streets are nodes of the graph.
Although intuitive and immediate, this type of representation is not unique.
For example, a circular square with a fountain in the middle can be represented as a closed chain as well as a single node. 
To each edge of the primary graph $G_p$, easily identifiable directly or by means of image processing algorithms \cite{URN}, we associate a label from a given set of urban properties. 

A suitable system of labels on the edges of $G_p$, together with the contraction process that will be defined, can make coherent different primary graphs representing the same space. 
The identification of the space syntax model is the identification of the labels measuring generic urban quantities.
Therefore we give the following definition:
\begin{definition}[Primary Graph $G_p$]
	A  primary graph $G_p$ (eventually georeferenced) is a pair of sets $(V,E)$ where $V$ is a set of geographic coordinates in $\mathbb{R}^2$ and $E\subseteq V\times V$.
	On $E$ is defined a function $l:E\rightarrow L$ which associate a label from a set $L$ to each $e\in E$.
	\label{gue}
\end{definition}
\ \\
In the following we discuss only non-directed graphs, assuming that in the ideal cities of the Renaissance there were no one-way streets or roundabouts.
In general $G_p$ is almost planar, in the sense that there may be a few edges corresponding to bridges or other overlapping structures which alter the planarity.
In this case two nodes with the same coordinates can be splitted using the coordinates in $\mathbb{R}^3$ instead of $\mathbb{R}^2$.
For simplicity we assume that all the $G_p$ coordinates are in $\mathbb{R}^2$.

\begin{figure}[!ht]
\centering
\includegraphics[width=0.6\textwidth]{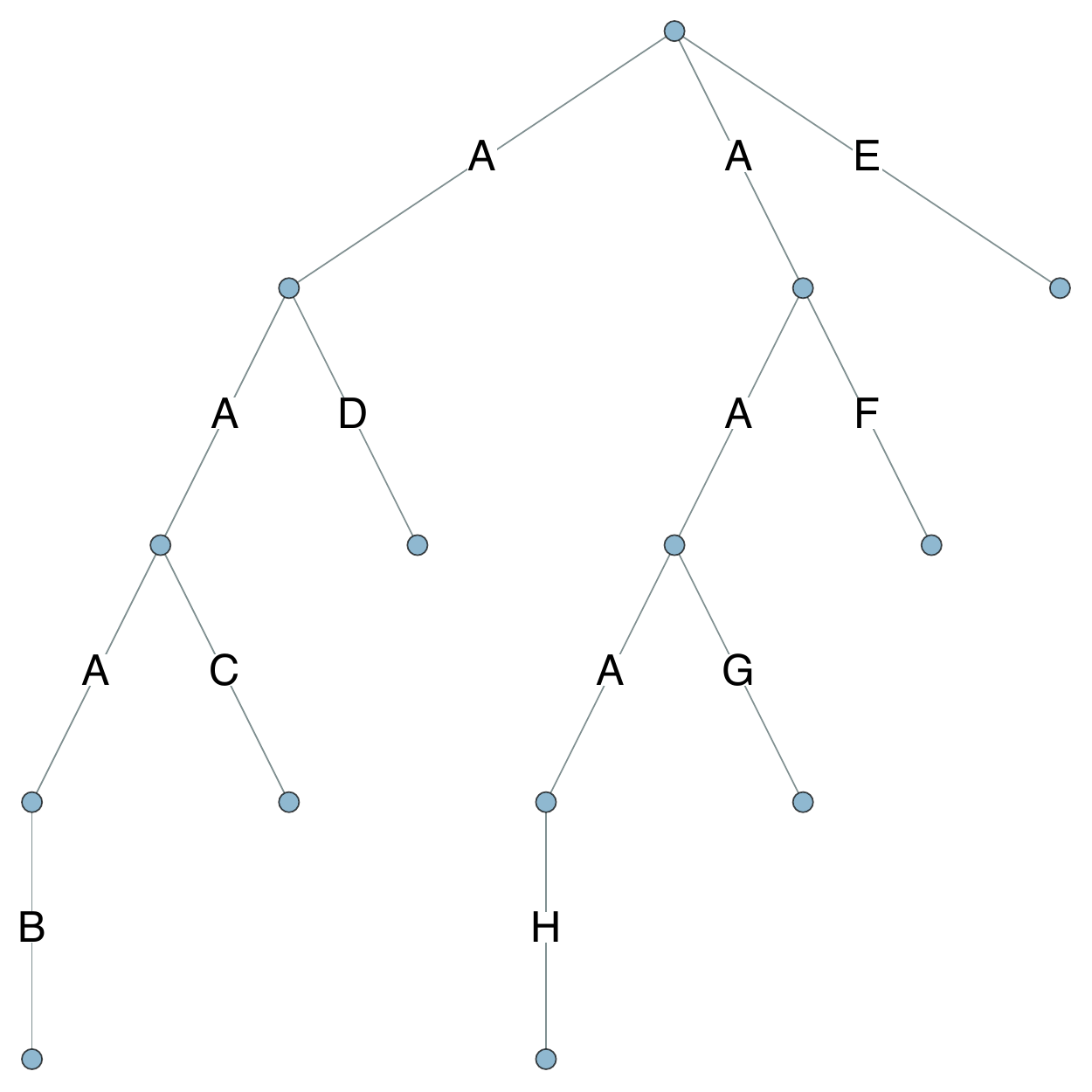}
\caption{The $G_p$ graph for an urban sprawl-like structure where the main street $A$ is connected to some dead-end streets $B,C,D,E,F,G,H$}
\label{fig4}
\end{figure}

To each edge of $G_p$ we associate a variable with values in a given set $C$ which measures an urban property.
We remark that this is not a graph colouring because adjacent edges can have the same color.
In the particular case of the space syntax, the labels represent the axial lines, but the same axial line can be broken up into connected edges that have the same label as shown in Fig. \ref{fig4}.

As mentioned above, the labels may indicate the street names, the type of road, the axial lines, the level of pollution or any other measurable urban quantity that may be associated with an open space.
To build the urban graph which gives as result the standard space syntax when adjacent labels identify the same line of view, we need the notion of {\sl line graph} \cite{Diestel2005}.\\
\begin{definition}[Line Graph $L(G_p)$]
	Given a graph $G=(V,E)$ the corresponding {\sl line graph} $L(G)$ is the graph on $E$ where two nodes of $L(G)$ are connected if the corresponding edges are adjacent in $G$.
\end{definition}
\ \\
It is evident that even when $G$ is a planar graph, the corresponding line graph will be not, in general, a planar graph.
The line graph drops all the coordinates of the nodes of $G_p$ but keeps all the information about the connectivity of the urban space.

For example a node of $G$ with degree $d$ becomes a {\sl clique} in $L(G)$ i.e. a subset where all the nodes are mutually connected.
Therefore, from the urban point of view, the line graph of $G_p$ is more interesting than the primary graph, because it shows a more complex structure.

\begin{figure}[!ht]
\centering
\includegraphics[width=0.6\textwidth]{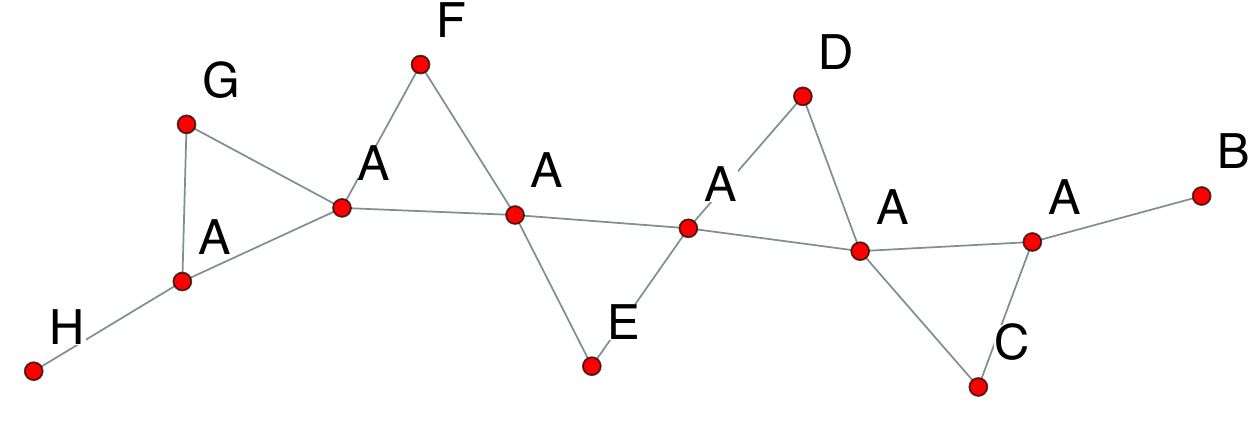}
\caption{The line graph $L(G_p)$ of the graph represented in Fig.\ref{fig4}}
\label{fig5}
\end{figure}

The labels of the $G_p$ edges are mapped on the corresponding nodes of $L(G_p)$.
Therefore $L(G_p)$ is a non-planar graph with labeled nodes.
This graph is not, apart from special cases, the same graph of the axial space syntax even when we label the axial lines.
For example in the case of the urban sprawl phenomenon (Fig. \ref{fig4}) the line graph does not return the star topology returned by space syntax \cite{VaroudisX}, but rather a graph where the main road $A$ is represented by many adjacent vertices with the same label (Fig. \ref{fig5}).

\begin{figure}[!ht]
\centering
\includegraphics[width=0.6\textwidth]{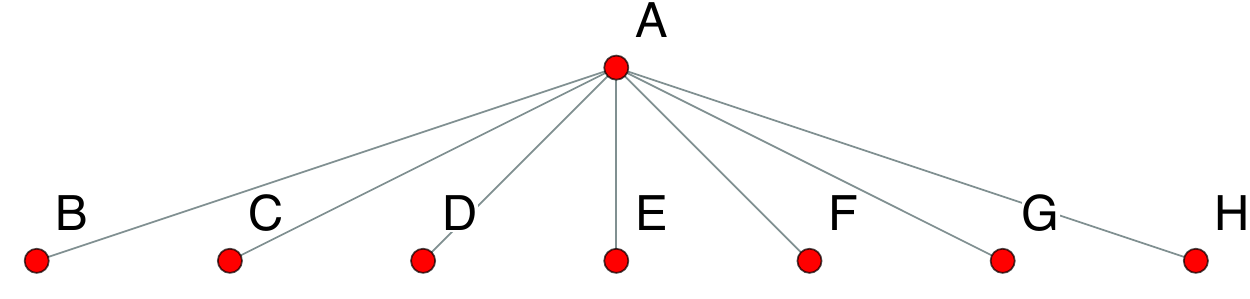}
\caption{The star-like graph given by the contraction of Fig.\ref{fig5}}
\label{fig6}
\end{figure}

\begin{figure}[!ht]
\centering
\includegraphics[width=0.6\textwidth]{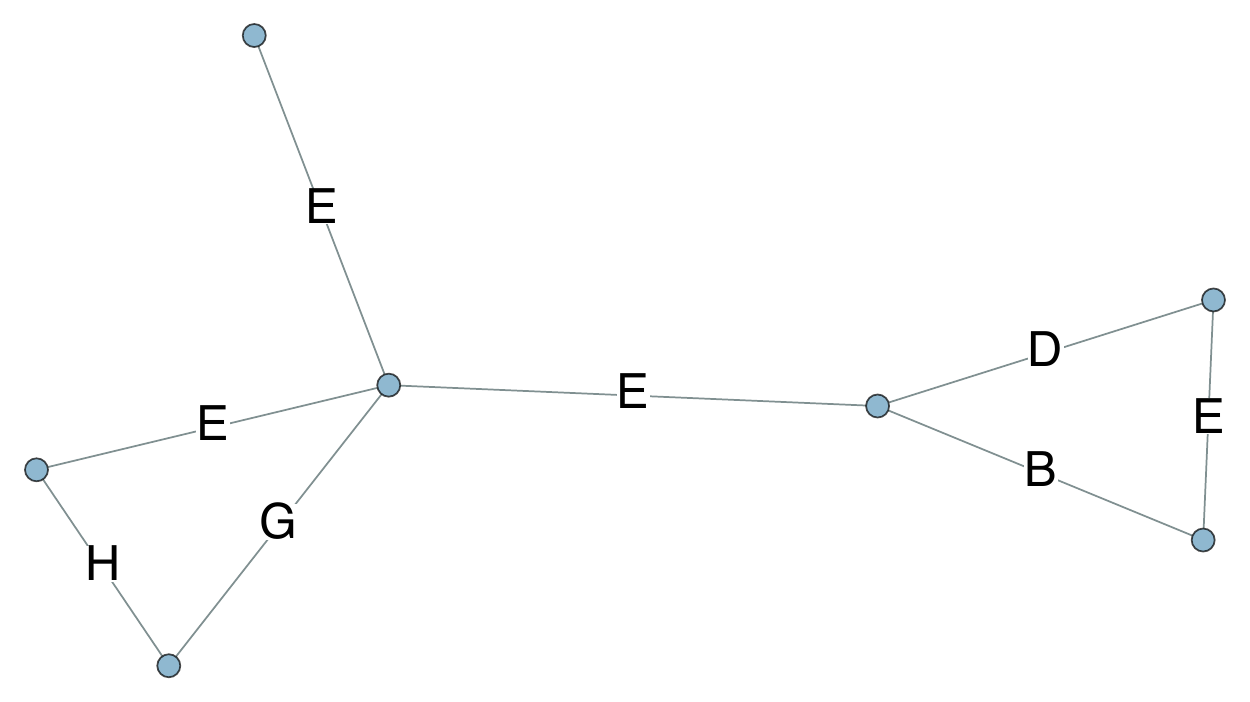}
\caption{The primary graph $G_p$ with labels on the edges}
\label{fig1}
\end{figure}

\begin{figure}[!ht]
\centering
\includegraphics[width=0.6\textwidth]{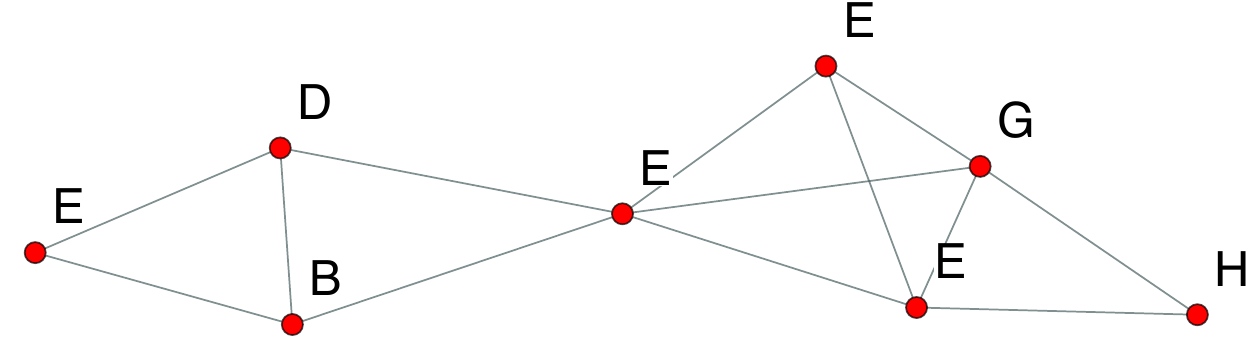}
\caption{The line graph $L(G_p)$ of Fig.\ref{fig1} before contraction}
\label{fig2}
\end{figure}

\begin{figure}[!ht]
\centering
\includegraphics[width=0.6\textwidth]{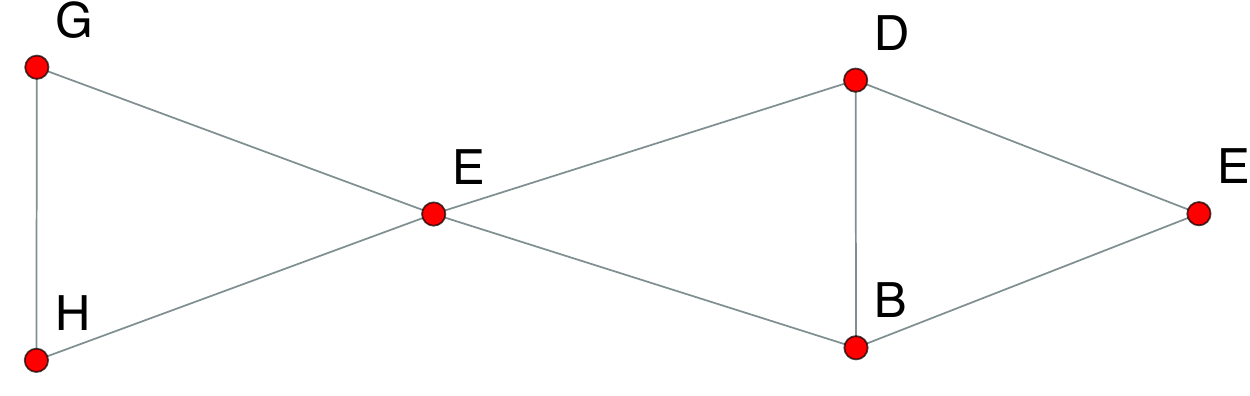}
\caption{The urban graph $G_U$ given by the contraction of Fig.\ref{fig2} }
\label{fig3}
\end{figure}

To complete the construction of the {\bf urban graph} it is necessary to introduce a suitable contraction procedure on the line graph.
The contraction \cite{Diestel2005} of the graph on one edge is the graph in which that edge (the pair of nodes) has been replaced by a new vertex which inherits all the edges of the two nodes removed.
Formally the contraction $G\setminus(x,y)$ of a graph $G$ on the edge $(x,y)$ is defined as the following:\\
\begin{definition}[Contraction]
	Given a graph $G=(V,E)$ and two vertices $x$ e $y$, the contracted graph on the edge $(x,y)$, $G_c=G\setminus(x,y)=(V_c,E_c)$ is the graph with the vertices $V_c=\big(V\setminus\{x,y\}\big)\cup v$ where $v$ is a new vertex, $v\notin V\cup E$, and $E_c= E\setminus(x,y)\cup\{(v,z)| (x,z)\in E \text{ or } (y,z)\in E\}$. 
\end{definition}\ \\
If we contract only the edges of $L(G_p)$ whose vertices have the same label, we are ready to give a definition of {\bf urban graph} $G_U$ which generalizes the space syntax: \\
\begin{definition}[Urban Graph $G_U$]
	Given a labelled line graph $L(G_p)$, the urban graph $G_U$ is obtained by the contraction of all the edges where the adjacent vertices have the same label.
	The new node $v$ given by the contraction of edge $(x_l, y_l)$ will have the same label $l$ of the two nodes.
	Multiple edges are transformed into a single edge.
\end{definition}
\ \\
When the labels correspond to the axial lines, the information on the urban spaces given by $G_U$ are consistent with the results of the space syntax.
For example, the contraction of Fig.\ref{fig5} gives a $G_U$ with a star topology shown in Fig.\ref{fig6}, which is the space syntax characterization of urban sprawl \cite{BV_Buch}.

In other words, we are grouping the open spaces in classes indexed by 1) the same value of the considered urban variable and 2) the adjacency of the corresponding line graph nodes.
The contraction and line graph operations do not commute: if we contract first on $G_p$ the adjacent edges with the same label and then we build the line graph, we get a different graph from $G_U$.
For the urban sprawl of Fig.\ref{fig1} the non-commutativity is evident.
We also observe that the interactions between the nodes of the contracted graph are still determined by the intersections.
However it is no longer the intersection of the axial lines that determine the interaction, but rather those of the generic spaces given by the equivalence classes of generic urban quantities.

We notice that while the construction of the spatial graph is a rigorous process, the set of labels for $G_p$ results from the choice of the urban quantities we want to measure.
The planners analyze different features of urban morphology (e.g. accessibility, axial lines, segmentation of public transport, noise, structural properties) and for each of these features measure different values of the labels on the different nodes of $G_p$.
However once this choice is made, the urban graph $G_U$ is given.

As explained in Sec.\ref{GU} our aim is to show the method of graph-construction starting from different points of view. Our set of invariant features could be easily replaced by other urban quantities or components, embracing all the spectrum of urban analysis, from singletons to whole-city features.
In the next section we analyze the $G_U$ properties for five ideal cities.
The set of labels is given by the structural elements $L=\{C,D,S,G,E,M,A_x\}$ identified in Sec.\ref{idealcity}.

\section{Results}
\label{results}
\begin{figure}[ht]
\centering
\includegraphics[width=160mm]{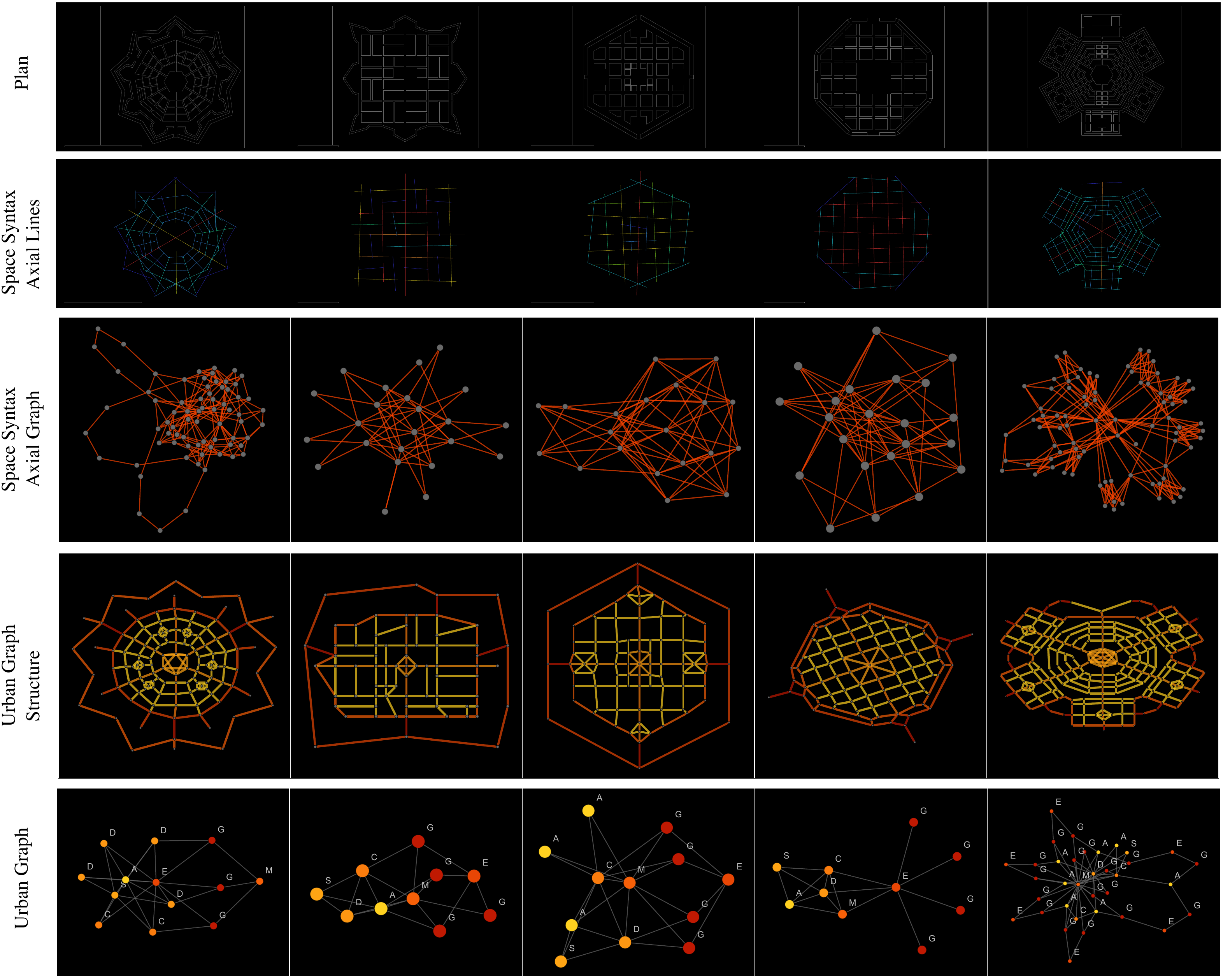}
\caption{From left the plans of Palmanova, Vitry le Fran\c{c}ois, Avola, Neuf Brisach, Grammichele and the corresponding space syntax and $G_U$ representations.}
\label{figAA}
\end{figure}

We analyze the urban structure of the five cities using both the space syntax and the $G_U$ given by the labels of tab.\ref{etichette}.
In fig.\ref{figAA} the plants of the cities are shown together with the axial lines, the space syntax graphs, the ``ideal'' structures and the corresponding $G_U$.
The $G_U$ has a smaller number of nodes compared to the space syntax graphs because the labels can assign the same value to adjacent nodes, for example the $A$ is a label for all the streets of the housing places.
The structural elements are represented by different colors (labels) as edges in the primary graph and nodes of $G_U$.
Some structural elements, for example the cardo and decumanus, are identical to those of space syntax showing that there is no incompatibility between the two representations.
Other spaces, such as the streets along the walls, do not have a corresponding element in space syntax, but also play an important role in connecting the city.

\begin{figure}[ht]
\centering
\includegraphics[width=160mm]{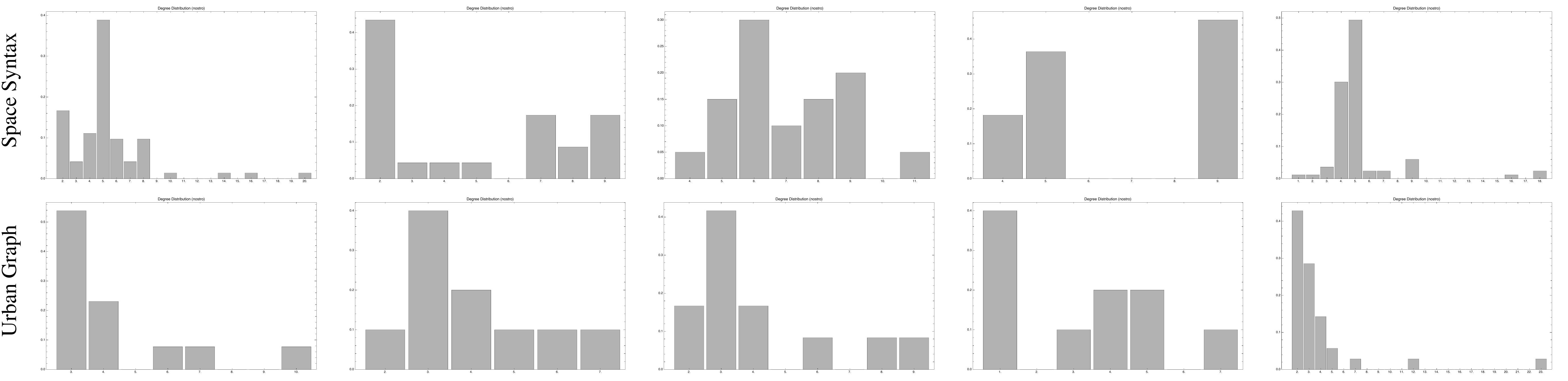}
\caption{The comparison of the Degree Distribution of the five cities for space syntax and $G_U$}
\label{figC}
\end{figure}

The $G_U$ and the axial graphs degree distributions are similar (fig.\ref{figC}).
In both the representations Palmanova and Grammichele show a node (the histogram tail) that connects many others spaces.
The central square serves as a link for both the axial lines and the structural elements of the ``ideal city'', especially when the city has a radial layout.
In the $G_U$ representation, the role of hub for the city it is not taken only by the cardo the decumanus or by the central square as in the case of the space syntax, but also by the external and internal streets along the walls.
The Urban Graph of Grammichele shows that the streets along the walls are the main connecting elements of the residential fabrics divided into blocks.
In the Urban Graphs of Vitry le Fran\c{c}ois, Avola and Neuf Brisach, the cardo and decumanus do not lose their connection function, but this function is weaker than that of the the streets along the walls.
This feature, which is not detected by the methods of the space syntax, is consistent with the military usage of the defense walls.

\begin{figure}[ht]
\centering
\includegraphics[width=160mm]{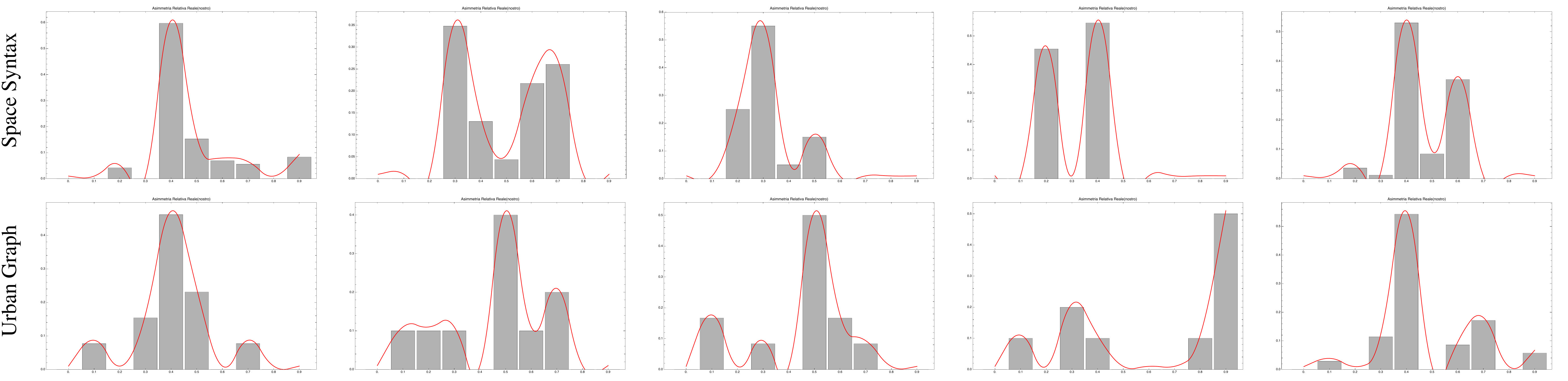}
\caption{The comparison of the Real Relative Asymmetry of the five cities for space syntax and $G_U$}
\label{figDD}
\end{figure}

Real Relative Asymmetry ($RRA$) gives a measure of the integration/segregation of the city.
For each node $i$ of the graph $G$ is defined as \cite{Kruger}
\begin{equation}
	RRA(G,i)={2(\mu(G,i)-1)\over(|G|-2)}D(|G|)^{-1}
\end{equation}
where
\begin{equation}
	\mu(G,i)=\sum_{j\in G}{d_{ij}\over|G|-1}
\end{equation}
$d_{ij}$ s the graph distance between the nodes $i$ and $j$, and $D(|G|)$ is 
\begin{equation}
	D(|G|)={2\big(|G|(\log_2{|G| + 2\over3} - 1) + 1\big)\over(|G|-2)(|G|-1)}
\end{equation}
In fig.\ref{figDD} the probabilities of the normalized values of $RRA$ and the smooth interpolation of the probability of the five cities are shown.
In a fully connected graph the distribution of $ARR$ probability is concentrated on the left and the city is well integrated.
A city with an organic structure has, in general, a bell-shape distribution for the $RRA$ probability \cite{BV_Buch}.

\begin{figure}[ht]
\centering
\includegraphics[width=160mm]{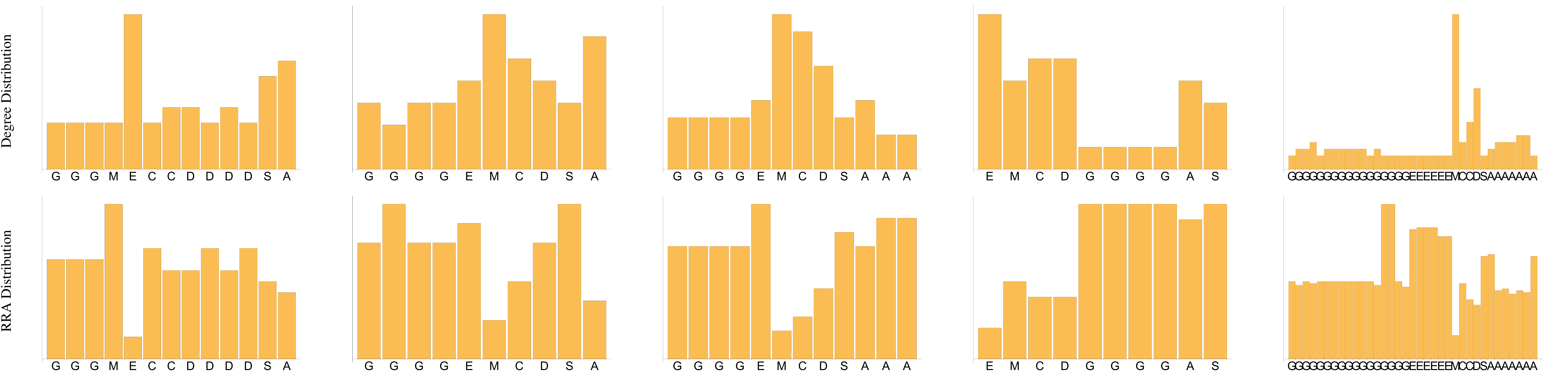}
\caption{$G_U$ Degree Distributiona and the Real Relative Asymmetry as function of the  urban structures}
\label{figEE}
\end{figure}

The Palmanova $RRA$ distribution shows this bell shape in both the representations, while the other cities show a more fragmented integration evidenced by more than one high probability peak.
The other cities are less integrated because the organic feature of Palmanova is due to its particular layout: the radial structure of the main streets induces a radial structure for the houses network.
Conversely the other cities show radial structure for the main roads, but a square lattice for the houses.
In the space syntax representation, the extreme case is given by Neuf Brisach, which shows two peaks, characteristic of a regular lattice structure with edge effects.
We observe that in the $G_U$ representation the cities are more integrated than in space syntax because once again the presence of the roads along the walls exerts a connecting function.
\\\\
The $G_U$ representation allows us to plot the degree distribution and the $RRA$ as functions of the structural urban elements.
For all the cities the degree of the nodes $E, M$ (the streets along the walls) is higher than that of the cardus and decumanus $C,D$.
Apart from the case of Neuf Brisach, the less segregated elements are the gates $G$, whereas the residential areas are the most segregated.
The graph of fig.\ref{figEE} shows that the military function, strongly influenced by the walls and the need to protect and defend the gates, seriously affects the organization of the city.
In fact inside a fortified city the inhabited areas should be the most protected, whereas the defense areas must be easily accessible from within the citadel.
Even though Grammichele was not built for military purposes, we see that this constraint induces the segregation of the residential areas.
However in this case the gates of the city are numerous, as are the connections for the many housing blocks.\\\\

Finally, to analyze the degree of connectivity of the city we compute the first non-zero eigenvalue of the normalized Laplace matrix, called spectral gap  $\lambda$ \cite{citeulike:667558}.
The normalized Laplace matrix ${\cal L}$ of a graph, related to the adjacency matrix, is defines as \cite{chungspectral}
\begin{equation}
	  {\cal L}(u,v)=\begin{cases}
	               1 \text{ if } u=v \qquad\text{ and } d_v\neq0\\
	               -1/\sqrt{d_ud_v}\qquad\text{ if } u,v \text{ are adjacent}\\
	               0\qquad\text{ otherwise}
	            \end{cases}	
\end{equation}
where $d_v$ is the degree of the vertex $v$.
If $\lambda_1,\ldots,\lambda_N$ are the eigenvalues of ${\cal L}$, if $\bar\lambda$ is the first non zero eigenvalue of ${\cal L}$ and $\lambda_N$ the last one (the biggest), we define the spectral gap $\lambda$ as $\lambda=\bar\lambda/\lambda_n$.
Therefore $\lambda$ shows how difficult it is to disconnect the city by closing or eliminating a few open spaces.
If $\lambda$ is small, the graph has a ``relatively clean bisection'', where a large $\lambda$ ``characterizes non-structured networks, with poor modular structure, in which a clearcut separation into subgraphs is not inherent'' \cite{citeulike:667558}.
The spectral gap of the five cities is small, according to the separation between center and periphery characteristic of the ancient cities.\\\\

The space-syntax-based axial graph suggests, indeed, how the two cities of Palmanova and Grammichele, those in which the radial structure is more emphasized, show a smaller value of $\lambda$. 
That means that they, in space syntax classical analysis, could be easily bisected, for instance making impossible the access to the central plaza.
In the formal representation of GU  based upon the process of labeling and contraction, in contrast, those cities are more difficult to cut across and reveal a greater structural integrity.
We see how although the $\lambda$ value of Grammichele, in the representation $G_U$,  is almost double the same value proposed by space syntax, it is still small compared with that of other cities, mainly because of the presence, along the radial roads, of urban clusters loosely connected to the wall system.

Finally, we should observe how, in the space syntax analysis, the city of Neuf Brisach seems to be very difficult to bisect, mostly due to the bipartite structure of the primary graph.
In a more accurate way, instead, and following the value extrapolated by the representation $G_U$,  are the cities of Vitry-le-Fran\c{c}ois and Avola the most difficult to split, due to the strong interconnection between different structural elements.
In this case we can observe the difference between a shape-based analysis, in which there is not a critique of the urban space, and a urban-based analysis, in which a space is labeled after its {\sl Innerste Wesen}, the intimate essence of a single place.

\begin{center}
	\begin{table}
		\centering
    \begin{tabular}{   c | c | c |}
    City & $\lambda$ (space syntax) & $\lambda(G_U)$ \\ 	[.1cm]\hline \\ [-.3cm]
 	Palmanova & 0.07276 & 0.23490 \\
  	Vitry le Fran\c{c}ois &  0.22269 & 0.23885 \\
  	Avola & 0.25967 & 0.27758 \\
  	Neuf Brisach &  0.31902 & 0.18663 \\
  	Grammichele & 0.05659 & 0.09782 \\
  	\end{tabular}
	\caption{The spectral gap of the ideal cities for the space syntax and $G_U$ representation.}
	\label{gap}
	\end{table}
\end{center}

\section{Conclusions}
\label{conclusions}
The main result of this study is that the main idea of space syntax, that is that the city is a relation network, could be easily extended also to non-spatial variables. Choosing these variables is the task of the urban planner; it is only after that these variables are identifyed that the process begins to be objective.
In the same way that for space syntax, the non-spatial urban values need to be related to the built environment.
The proposed method is built upon two pillars, the first is the more accurate definition of line graph, and the second one is the actions of labeling and contraction.

The suggested method, if the labeling and contraction phase is applied to the axial lines, as theorized by space syntax, returns the same measures.
In this sense we can affirm that space syntax theory of axial lines is a peculiar configuration of the proposed method.
The analysis of the urban graph $G_U$ of the ideal cities returns results that are different, if compared to space syntax, but more coherent from an urban point of view.
As we stressed in Sec.\ref{GU}, i.e. in the case of the sprawl city.

In particular we could observe how the presence of roads along the walls represent an improving value of connectivity of the built environment, whereas in space syntax, these roads are only another part of the network, therefore liable to edge errors which in our method are largely limited.

The analysis of the spatial graphs of ideal cities leads to several considerations.
The method of label contraction turns out to be a useful tool to identify and measure the characteristics of urban structures related to flows and accessibility.
However it depends on the system of primary graph edge labels representing the variables we want to take into account.
The properties of the spatial graphs show that the geometric structure is more efficient when it is amended by a reticular or even disordered structure which increases its robustness.

The bottlenecks represented by the city gates can be measured on the spatial graph, to assess the effectiveness of a fortification, by means of the level of difficulty required to enter or exit from the same city.
In all the cases considered the  difference between center and periphery is clear, although the roads adjacent to the walls, while geographically peripheral, play a fundamental connection function.

The Renaissance town is at the basis of the ``modern'' city and the ideals of this city are identified in the principles of rationality and equality represented by a geometric order.
This consideration is partly revised by the results of our analyses because one more disordered development within this geometric order strengthens stability and accessibility, at least in terms of graph theory.

The shape of the city also expresses the change of a political vision: the shape of New York, orthogonal and rectangular, abhors the boulevard structure of the cities of monarchic Europe and establishes a criterion of cost-effectiveness. In other words it opposes to one old-fashioned the ideal city a different idea of ideal city.
In the paper we analyzed the Western fortified cities of the sixteenth and seventeenth centuries, but we must necessarily broaden the discussion to other ideas of the city, which may be at odds with one another.
In a further work we will try to apply the same method to an idea of urban relationship based on acoustics.
It is therefore to be imagined how different forms of civilization generate different taxonomies and labeling, and whether they can be applied in New York as well as in in Arbil, and if, finally, the civilization of the ideal city is still able to produce meanings that a planner can recognize, isolate, evaluate, label and correlate to life among the buildings.
The authors would like to thank Paul Blanchard and Paolo Desideri for their valuable comments which helped to improve this research.

\bibliographystyle{plain}
\bibliography{/Users/dautilia-n/Desktop/DTC/WorkInProgress/References/bibliografia_rob} 
\end{document}